\documentclass[preprint,12pt,authoryear]{elsarticle}

\usepackage{amssymb}
\usepackage[version=4]{mhchem}

\usepackage{multirow}
\usepackage{color, soul}
\usepackage{booktabs}
\usepackage{subcaption}
\journal{ NIMB: Beam Interactions with Materials and Atoms}

\begin{document}

\begin{frontmatter}



\title{Influence of DFT Functionals on Low-Energy Electron Scattering Cross Sections of Nitric Oxide}


\author[inst1]{Ashutosh Yadav}

\affiliation[inst1]{organization={Department of Physics, Indian Institute of Technology (Indian School of Mines)},
            city={Dhanbad},
            postcode={826004}, 
            state={Jharkhand},
            country={India}}
\author[inst2,inst4]{Felipe Fantuzzi}
\affiliation[inst2]{organization={Chemistry and Forensic Science, School of Natural Sciences, University of Kent},
           addressline={Park Wood Rd}, 
            city={Canterbury},
            postcode={CT2 7NH}, 
            country={United Kingdom}}

\author[inst3,inst4]{Nigel J. Mason}
\affiliation[inst3]{organization={Physics and Astronomy, School of Engineering, Mathematics and Physics, University of Kent},
           addressline={Park Wood Rd}, 
            city={Canterbury},
            postcode={CT2 7NH}, 
            country={United Kingdom}}

\affiliation[inst4]{organization={HUN-REN Institute for Nuclear Research (Atomki)},
            city={Debrecen},
            postcode={H-4026},
            country={Hungary}}

\author[inst1,inst3]{Bobby Antony}
\ead{bobby@iitism.ac.in}

\begin{abstract}
Nitric oxide (NO) is important in biological, atmospheric, plasma, industrial, and astrophysical environments, where reliable electron-collision data support modelling charged-particle interactions with matter. Its well-known experimental properties make it suitable for assessing how the target electronic-structure description affects low-energy electron scattering calculations. In this work, NO properties were evaluated using B3LYP, M06-2X, PBE0, and $\omega$B97X-D3, with basis sets ranging from minimal to quadruple-zeta quality. Bond length, dipole moment, ionisation potential, and polarisability were compared with experiment to assess the sensitivity of the target description to the functional and basis set. The aug-cc-pVQZ basis set was then used to generate target models for ab initio R-matrix calculations over 0.1--20 eV. The total cross sections show low-energy resonance features, with the strongest functional dependence around the broad peak near 0.8--1.0 eV. A sharper, higher-energy structure is also observed below 2 eV, shifting from 1.74 to 1.82 eV depending on the functional. Differential cross sections show modest functional sensitivity, with more noticeable angular differences at 7.5 and 10 eV. These results show that the DFT functional and basis set affect the target properties, with the resulting target description influencing low-energy electron-scattering observables of NO. The comparison supports $\omega$B97X-D3/aug-cc-pVTZ geometry optimisation followed by aug-cc-pVQZ target-property calculations as a practical protocol for R-matrix modelling of NO.
\end{abstract}
 
\begin{graphicalabstract}
\includegraphics[scale=0.55]{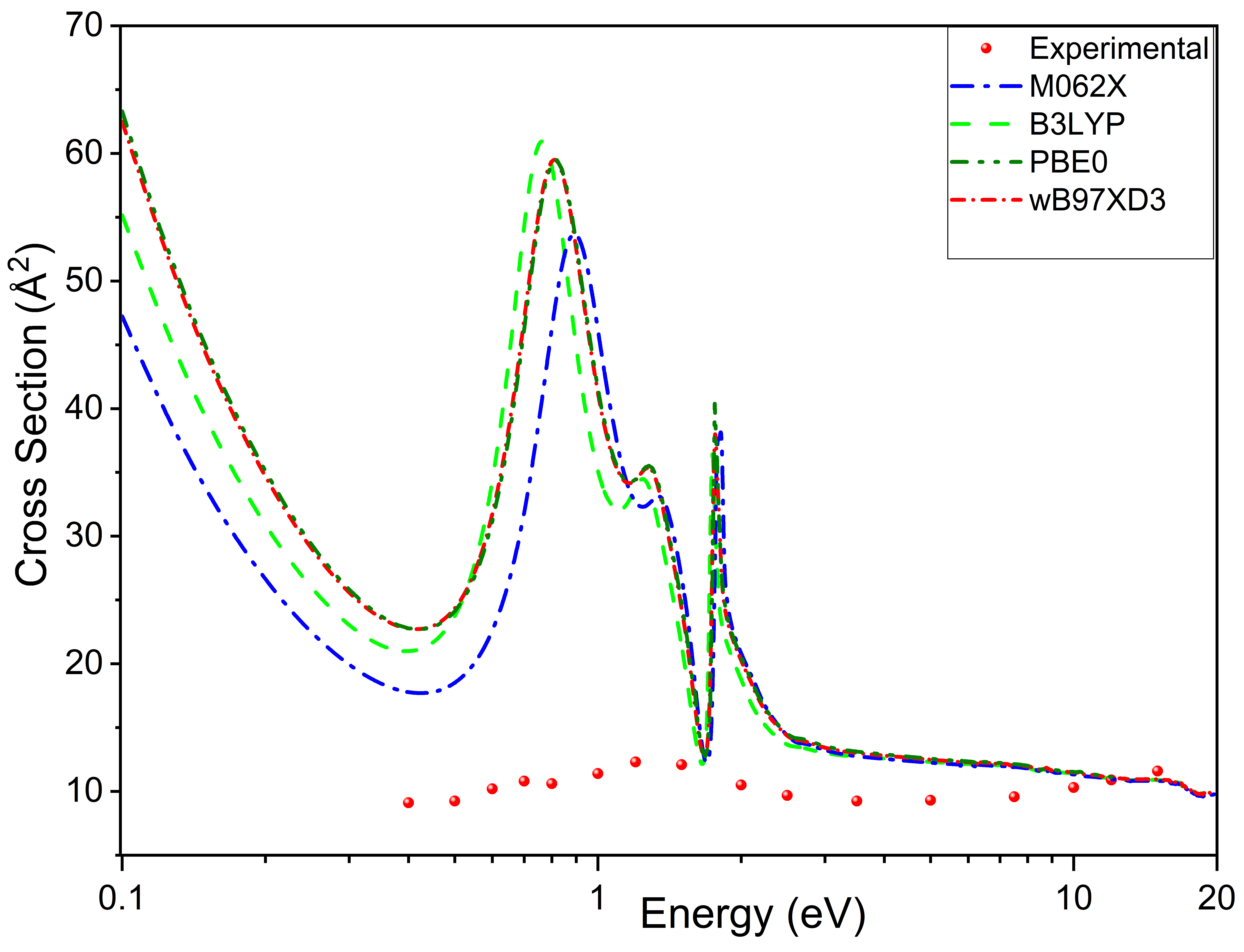}
\end{graphicalabstract}

\begin{highlights}

\item DFT-based target descriptions influence low-energy electron scattering from nitric oxide.
\item Molecular properties of nitric oxide show distinct functional and basis-set sensitivities.
\item Total cross sections are strongly affected by functional choice in the resonance region.
\item Differential cross sections show weaker functional dependence than total cross sections.
\item $\omega$B97X-D3/aug-cc-pVTZ geometries with aug-cc-pVQZ target properties provide a balanced R-matrix protocol.

\end{highlights}


\end{frontmatter}




\section{Introduction}

Low-energy electrons are ubiquitous in molecular environments, arising from ionising radiation and secondary-electron cascades \citep{Pimblott2007,Alizadeh2015}, as well as from electrical discharges and plasma processes \citep{Bogaerts2002}. Although their kinetic energies are often modest, low-energy electrons can interact efficiently with molecules through elastic scattering, excitation, ionisation, and dissociative processes. These collision pathways influence energy deposition, electron transport, molecular fragmentation, and reactive-species formation, thereby affecting the energy distribution and chemical composition of gases, plasmas, planetary atmospheres, biological media, irradiated materials, and astrophysical ices \citep{Brunger2017,Campbell2016,Alizadeh2015,Sanche2003,Dickers2025,Mason2026,Zhang2024}. Electron-scattering cross sections are therefore required as quantitative input for modelling electron transport, molecular fragmentation, and radiation-induced chemistry, and their accuracy is essential for reliable predictions \citep{ARYA2, NOREF,DEANO}.

At low incident energies, typically below 15 eV, resonance scattering can make a substantial contribution to the elastic and total cross sections. In this regime, temporary attachment of the incoming electron to the molecular target may enhance scattering probabilities and create efficient channels for excitation and dissociation. The ab initio R-matrix method \citep{HT,PM} is widely used in electron--molecule collision studies because it provides a reliable description of resonance positions and widths, together with the associated scattering observables.

The quality of the calculated cross sections depends not only on the scattering model, but also on the molecular description of the target. Properties such as the ionisation potential, equilibrium bond length, polarisability, and dipole moment can affect the interaction between the incident electron and the molecule, especially in the low-energy regime \citep{NAGHMA201317,ARYA2,AY2025}. In density functional theory (DFT) calculations, these properties are sensitive to the choice of exchange--correlation functional and basis set, meaning that different electronic-structure descriptions may lead to different scattering results. Assessing this sensitivity is therefore important for selecting reliable target models before performing cross-section calculations \citep{30YEARSDFT}.

In this work, we evaluate the dependence of the molecular properties of nitric oxide (NO) on the choice of density functional and basis set. NO is a suitable target for this purpose because its ionisation potential, bond length, polarisability, and dipole moment are experimentally well characterised \citep{CCCBDB}, and reference electron-scattering cross sections are available over the 0.1--20 eV energy range \citep{NOREF}. As a small open-shell diatomic molecule with an unpaired electron in an antibonding $\pi^\ast$ orbital, NO also has distinctive electronic, spectroscopic, and scattering properties \citep{Shiotari2021}. It is relevant in atmospheric, plasma, combustion, biological, and astrophysical environments \citep{Liszt1978,Barth2003,Krasnopolsky2006,Stoffels2006,Hughes2008,Liu2025}, where its interactions with electrons contribute to excitation, ionisation, and dissociation processes \citep{RES,NOREF}. The molecular properties were calculated using basis sets ranging from minimal descriptions, such as STO-3G and 3-21G, to Pople, Dunning, and Karlsruhe basis sets of increasing flexibility. Four density functionals were considered: B3LYP \citep{B3LYP} and PBE0 \citep{PBE0,Ernzerhof1999}, both hybrid GGA functionals with 20\% and 25\% Hartree--Fock (HF) exchange, respectively; M06-2X \citep{M06}, a Minnesota meta-hybrid GGA functional with 54\% HF exchange; and $\omega$B97X-D3 \citep{Lin2013}, a range-separated hybrid functional with dispersion correction and long-range HF exchange increasing to 100\%.

After benchmarking the molecular properties against experimental data, the aug-cc-pVQZ basis set was selected for the scattering calculations because it provides an accurate description of the target electronic structure, including adequate radial and angular flexibility and diffuse character. Electron-scattering cross sections were then calculated using the R-matrix method \citep{tennyson1,tennyson2} with target models generated from the different density functionals. This protocol allows the sensitivity of total and differential cross sections to the underlying functional choice to be assessed while keeping the basis set fixed.

Section 2 describes the theoretical methodology, including the R-matrix approach and the DFT calculations used to determine the molecular properties. Section 3 presents the calculated molecular properties obtained with different basis sets and density functionals, followed by the analysis of total and differential cross sections. Section 4 summarises the main conclusions, with emphasis on the effect of basis-set and exchange--correlation functional choices on the calculated scattering observables.

\section{Theory}

The low-energy electron cross-sections were calculated using the R-matrix method, with molecular properties obtained from DFT calculations using the ORCA software \citep{neese2020orca}. The molecular properties, such as ionisation energy, bond length, polarisability, and dipole moment, are calculated by optimising the NO structure using different basis sets and four functionals: B3LYP, M06-2X, PBE0, and $\omega$B97X-D3 to see the effect of exchange and correlation on the molecular properties. The cross-section calculations were performed using the R-matrix method \citep{tennyson1}, as implemented in the Quantemol-N program \citep{tennyson2}. The R-matrix approach is a well-established framework for electron-molecule collision studies. It works by dividing configuration space into two regions separated by a spherical boundary of radius \(a\) defined with respect to the centre of mass of the target molecule. The inner region contains all $N$ electrons of the target molecule together with the incoming projectile electron, forming a total \((N+1)\)-electron system. In this region, short-range interactions such as electron exchange and correlation are dominant, and the \((N+1)\)-electron system effectively behaves as a bound system over short distances. In the outer region, short-range interactions such as exchange and correlation no longer influence the scattered electron, and the interaction is determined by the long-range multipole potential of the target molecule. The scattering wavefunction in the outer region is expressed in a form that leads to a set of coupled differential equations, which are then solved using standard numerical techniques.

To ensure that short-range interactions vanish at the boundary, the radius of the R-matrix sphere must be chosen carefully, typically in the range 10--15 $a_0$, so that the \((N+1)\)-electron charge density is contained within it. In the present work, a radius of 12 $a_0$ was used. Continuity of the wavefunction is ensured by matching the solutions in the inner and outer regions at the boundary, although the two regions are treated independently.

The wavefunction in the inner region is represented using the close-coupling approximation:
\begin{equation}
\psi^{N+1}_k
= \hat{A} \sum_{i j} a_{i j k}\, \Phi_i(x_1, x_2, \ldots, x_N)\, u_{i j}(x_{N+1})
\;+\; \sum_i b_{i k}\, \chi_i(x_1, x_2, \ldots, x_{N+1}),
\end{equation}
where \(\hat{A}\) is the antisymmetrisation operator. In this expression, \(\Phi_i\) are the wavefunctions of the target molecule, while \(u_{i j}\) are continuum orbitals representing the scattering electron. The coefficients \(a_{i j k}\) give the weight of each target–continuum combination in forming the inner-region solution. The terms \(\chi_i\) are square-integrable (\(L^2\)) functions confined within a finite region, and the coefficients \(b_{i k}\) describe their contribution to the solution.

The R-matrix is constructed at the boundary from the inner-region eigenfunctions and is then propagated outward and compared to asymptotic solutions obtained using the Gailitis expansion \citep{gailitis}. Through this procedure, the \(K\)-matrix is determined. Using the POLYDCS code \citep{sanna1998differential}, the \(K\)-matrix is converted into the \(S\)- and \(T\)-matrices, which are directly related to the scattering observables. The scattering cross sections are then
calculated from the elements of the \(T\)-matrix. In the present study, a configuration interaction (CI) model was employed, providing a highly accurate description within the present R-matrix framework by including static, exchange, and polarisation effects, as well as target excitations required to capture resonance effects and describe excitation dynamics reliably.

\subsection{Target State Modelling}

Nitric oxide has \(C_{\infty v}\) point group symmetry. However, since this symmetry cannot be implemented in the present model, the reduced point group $C_{2v}$ was used for the target, and the scattering cross-sections were calculated using molecular properties obtained from different functionals in the aug-cc-pVQZ basis set. A CI model was employed for the scattering calculations. The ground-state electronic configuration of NO is:  
\[
1(a_1)^2,\; 2(a_1)^2,\; 3(a_1)^2,\; 4(a_1)^2,\;
1(b_1)^2,\; 5(a_1)^2,\; 1(b_2)^2,\; 2(b_2)^1
\]

\noindent Out of the 15 electrons, the innermost 8 electrons were frozen in the $(1a_1,\;\allowbreak 2a_1,\;\allowbreak 3a_1,\;\allowbreak 4a_1)^{8}$ orbitals, and the remaining 7 electrons were treated as active and distributed over 9 orbitals, $(5a_1,\;\allowbreak 6a_1,\;\allowbreak 7a_1,\;\allowbreak 1b_1,\;\allowbreak 2b_1,\;\allowbreak 3b_1,\;\allowbreak 1b_2,\;\allowbreak 2b_2,\;\allowbreak 3b_2)^{7}$. In the CI calculation, a total of 1524 configuration state functions (CSFs) were generated for 18 target states.

\section{Results and Discussion}

Calculated bond lengths, dipole moments, ionisation potentials, and polarisability are compared in Figures 1 to 4 for different basis sets, ranging from STO-3G to quadruple-zeta quality using four different type of functional B3LYP, M06-2X, PBEO, $\omega$B97X-D3 which differ by amount of excahnge, and for that we have used the qunatum chemical software ORCA\cite{neese2020orca}. After calculating and comparing these properties with experiment, the electron-impact total cross sections are obtained for the energy range 0.1--20 eV. 

\subsection{Bond Length}

\begin{figure}[ht]
\centering
  \includegraphics[scale=0.395]{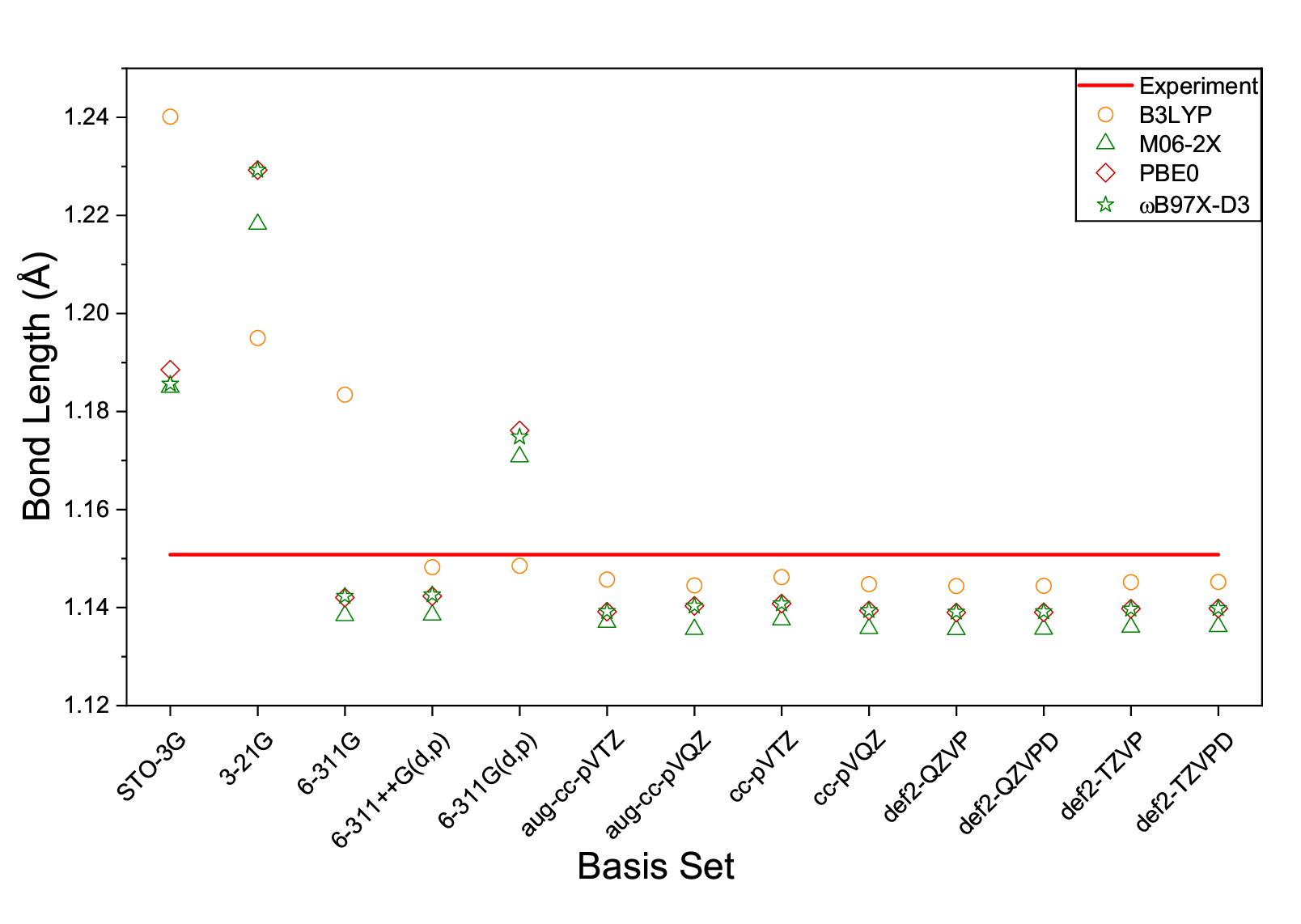}
\caption{Calculated equilibrium bond lengths of NO as a function of basis set for different density functionals. The experimental bond length is indicated by a horizontal reference line.}
  \label{fig1}
\end{figure}

As shown in Figure \ref{fig1}, the calculated bond lengths are strongly dependent on the choice of basis set. Minimal basis sets such as STO-3G and 3-21G systematically overestimate the equilibrium distance for all functionals, reflecting their limited ability to describe electron-density redistribution and orbital polarisation during bond formation. The inclusion of polarisation and diffuse functions, together with increased flexibility at the triple-zeta level, significantly improves agreement with experiment. In this regime, the 6-311++G(d,p) and 6-311G(d,p) basis sets provide the most accurate results across the functionals considered. Among the functionals, B3LYP yields the closest agreement with experiment for these basis sets, while PBE0 and $\omega$B97X-D3 show slightly larger but comparable deviations. The largest discrepancies are observed for M06-2X, which tends to produce shorter equilibrium distances. This behaviour is consistent with the relatively high fraction of HF exchange (54\%) in M06-2X, which has been shown to introduce systematic deviations in equilibrium geometries for main-group systems in benchmark studies \citep{Zhao2008}, with a tendency to produce shorter bond lengths in specific cases \citep{Huh2010}. The percentage deviation in bond length from the experimental value is largest for the STO-3G basis set with the B3LYP functional, followed by the 6-311++G(d,p) basis set. The smallest deviation, approximately $0.20\%$, is obtained with the 6-311G(d,p) basis set. For larger basis sets, the deviation obtained using B3LYP generally ranges from $0.20\%$ to $0.60\%$, whereas the other functionals exhibit deviations of approximately $1$--$3\%$ with triple- and quadruple-zeta basis sets, with larger deviations generally observed for smaller basis sets.

\subsection{Dipole Moment}

\begin{figure}[ht]
\centering
  \includegraphics[scale=0.395]{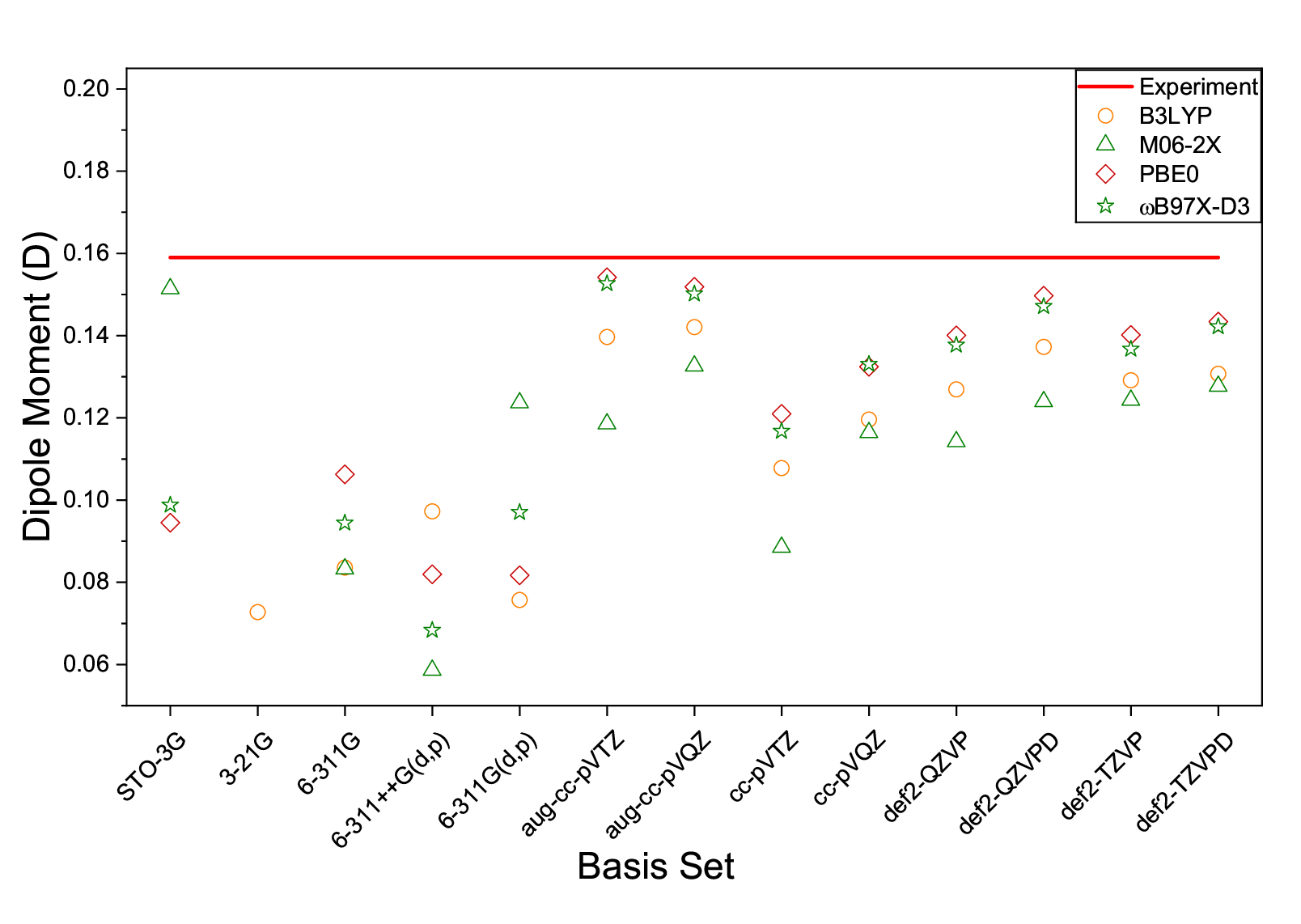}
  \caption{Calculated dipole moments of NO as a function of basis set for different density functionals. The experimental dipole moment is indicated by a horizontal reference line.}
  \label{fig2}
\end{figure}

The calculated dipole moments are presented in Figure \ref{fig2}, which shows that all results underestimate the experimental value. The aug-cc-pVQZ and aug-cc-pVTZ basis sets give dipole moments closest to experiment when combined with the PBE0 and $\omega$B97X-D3 functionals, highlighting the importance of diffuse functions in accurately describing the electron density and charge distribution in NO. The improved performance of these basis sets, compared to the def2 (Ahlrichs-type) basis sets, reflects the correlation-consistent design of the aug-cc basis sets, which ensures systematic convergence and a more reliable description of the diffuse electron density, particularly in the asymptotic (long-range) region \citep{Kirschner2020}. The def2 basis sets, including def2-QZVPD, def2-TZVPD, def2-TZVP, and def2-QZVP, also include diffuse functions but are less accurate overall.

The B3LYP and M06-2X functionals yield slightly lower dipole moments than PBE0 and $\omega$B97X-D3 across all basis sets considered. Furthermore, cc-pVTZ, cc-pVQZ, 6-311++G(d,p), and 6-311G(d,p) basis sets produce dipole moments that are significantly lower than the experimental value for all functionals. In the 3-21G basis set, M06-2X gives a dipole moment close to experiment; however, this agreement appears fortuitous, as all other basis sets show larger deviations.

In general, PBE0 and $\omega$B97X-D3 provide the best agreement with experiment. This can be attributed to their more balanced treatment of exchange and correlation, which yields a more accurate electron density distribution and charge separation in NO. In contrast, B3LYP, with a lower fraction of exact exchange, and M06-2X, with a significantly higher fraction (54\%), tend to underestimate the dipole moment: the former due to residual self-interaction and associated delocalisation errors arising from the incomplete cancellation of the self-Coulomb interaction, which can lead to an over-delocalised electron density \citep{Bao2018}, and the latter due to increased electron localisation associated with high HF exchange. The percentage deviation in the dipole moment from the experimental value for the triple- and quadruple-zeta basis sets ranges from approximately $4\%$ with the aug-cc-pVTZ basis set to $40\%$ with the 6-311++G(d,p) basis set for the $\omega$B97X-D3 functional. The PBE0 and B3LYP functionals exhibit similar variations in deviation across these basis sets. The maximum deviation of $47.64\%$ is seen for the M06-2X functional using the 6-311++G(d,p) basis set, indicating a substantial discrepancy from the experimental dipole moment.  

\subsection{Ionisation Potential}
\begin{figure}[ht]
\centering
  \includegraphics[scale=0.395]{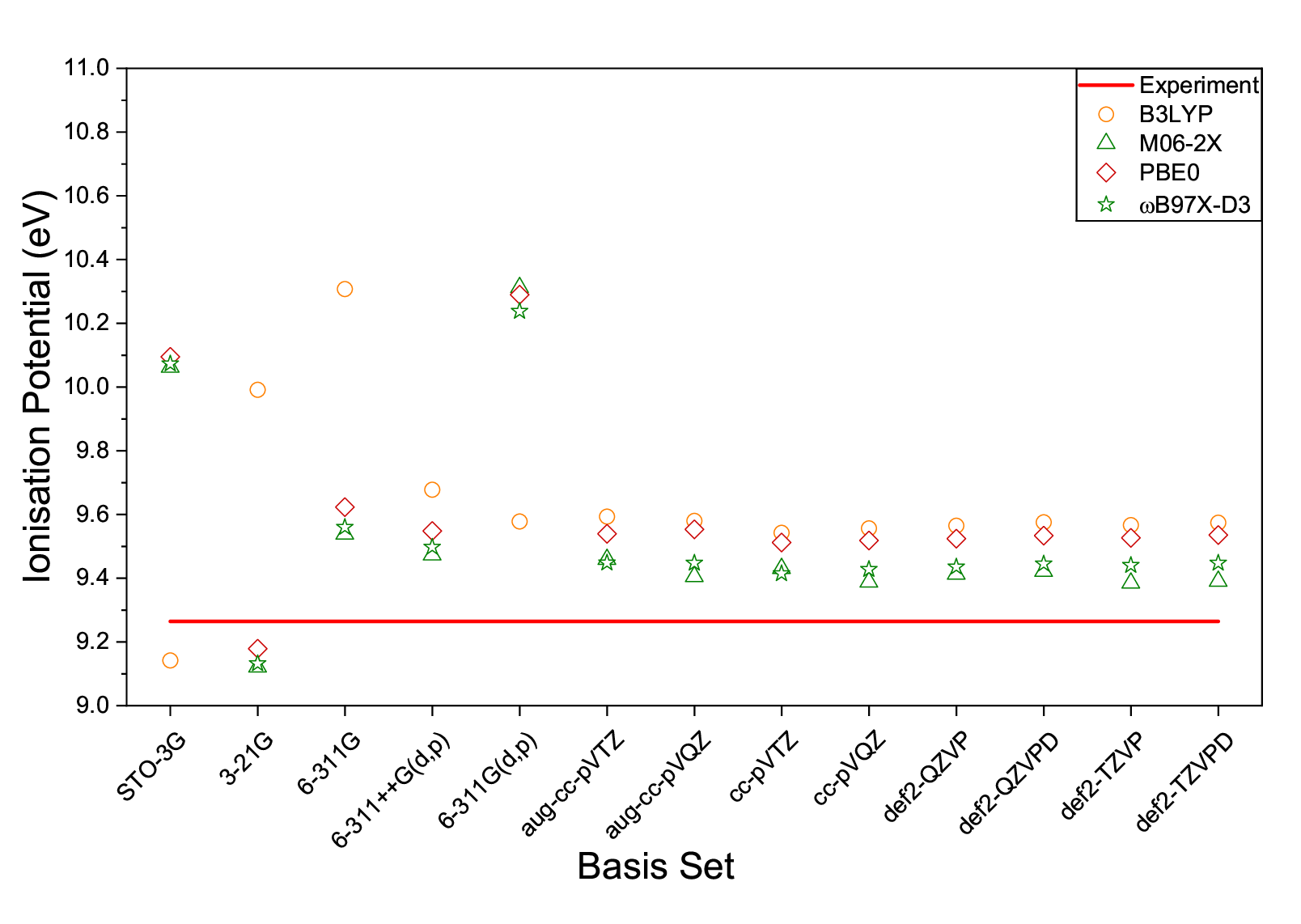}
  \caption{Calculated ionisation potentials of NO as a function of basis set for different density functionals. The experimental ionisation potential is indicated by a horizontal reference line.}
  \label{fig3}
\end{figure}

The ionisation potentials shown in Figure \ref{fig3} indicate that the calculated values generally overestimate the experimental reference for most basis sets, with the exception of the minimal STO-3G and 3-21G cases, which display erratic behaviour. For small and intermediate basis sets, such as STO-3G, 3-21G, and 6-311G, the results vary significantly across functionals, reflecting an insufficient and inconsistent description of the electronic structure.

From the triple-zeta level onwards, particularly for aug-cc-pVTZ and larger basis sets, the calculated ionisation potentials converge and become largely insensitive to further basis-set improvement. In this regime, the results cluster according to the choice of functional, indicating that the remaining deviations are primarily functional-driven rather than basis-set limited.

Among the functionals, a clear trend is observed: B3LYP yields the largest overestimation of the ionisation potential, followed by PBE0, while $\omega$B97X-D3 and M06-2X provide values closer to experiment. The improved performance of M06-2X and $\omega$B97X-D3 can be attributed to their higher effective fraction of HF exchange, which improves the description of electron removal and partially mitigates self-interaction error. In contrast, B3LYP and PBE0, with lower fractions of exact exchange, tend to overestimate the ionisation potential more significantly. The percentage deviation in the ionisation potential is relatively small for the triple- and quadruple-zeta basis sets, generally ranging from $1\%$ to $3\%$ for the 6-311G(d,p) basis set. The maximum deviation across the different functionals is approximately $5\%$, indicating generally good agreement with the experimental ionisation potential.

Overall, the results highlight that, once a sufficiently flexible basis set is employed, the accuracy of the ionisation potential is predominantly determined by the exchange--correlation functional.

\subsection{Polarisability} 

\begin{figure}[ht]
\centering
  \includegraphics[scale=0.395]{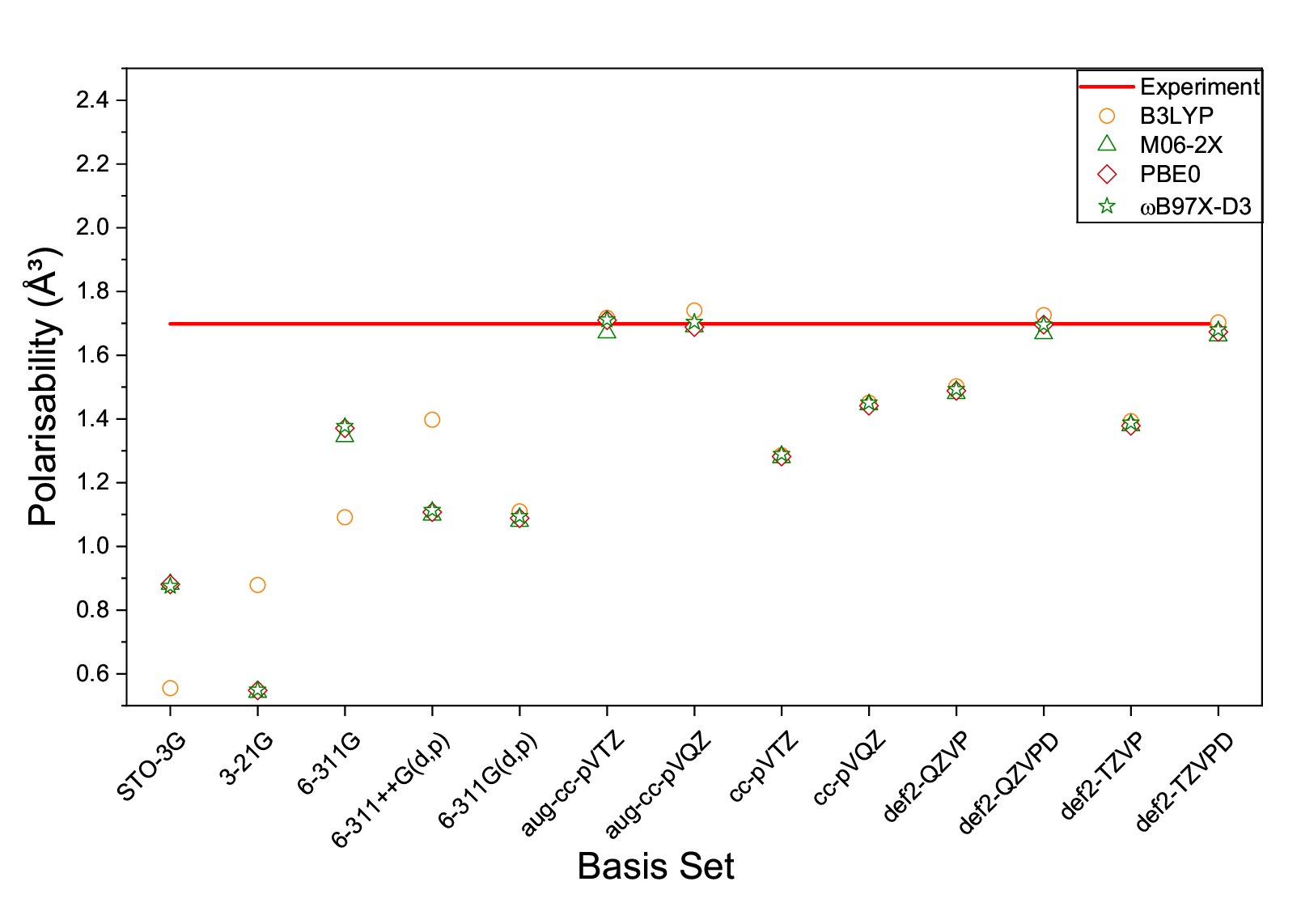}
  \caption{Calculated polarisabilities of NO as a function of basis set for different density functionals. The experimental polarisability is indicated by a horizontal reference line.}
  \label{fig4}
\end{figure}

The calculated polarisability of the NO molecule is presented in Figure \ref{fig4}. The results show that polarisability is strongly dependent on the choice of basis set, while the influence of the density functional is comparatively weak, with all functionals yielding similar values across the basis sets considered. Minimal basis sets such as STO-3G and 3-21G severely underestimate the polarisability, reflecting their inability to describe the distortion of the electron density in response to an external field.

A significant improvement is observed with the inclusion of polarisation and, more importantly, diffuse functions, as seen for the 6-311G-based and aug-cc basis sets. The most accurate polarisability values are obtained with the aug-cc-pVTZ and aug-cc-pVQZ basis sets, as well as with def2-QZVPD and def2-TZVPD, which provide results closest to experiment. This highlights the critical role of diffuse functions in capturing the asymptotic response of the electron density. In contrast, non-augmented basis sets such as cc-pVTZ and cc-pVQZ systematically underestimate the polarisability. The percentage deviations in polarizability from the experimental values are relatively small, ranging from approximately $0.2\%$ to $2\%$ for all functionals with triple- and quadruple-zeta basis sets. In contrast, the Pople-type basis sets generally exhibit larger deviations. The deviations obtained with different functionals are relatively small, with the exception of the B3LYP functional, which shows comparatively larger deviations in polarizability.

The improved performance of triple- and quadruple-zeta basis sets reflects their increased flexibility in the valence region, combined with a more appropriate description of diffuse electron density. These results demonstrate that an accurate representation of the electron-density tail is essential for reliable polarisability calculations, and that basis-set quality is the dominant factor determining the accuracy of this property.

The molecular-property analysis shows that no single functional and basis-set combination gives the closest agreement with experiment for all target properties. The bond length is best reproduced with B3LYP in combination with 6-311++G(d,p) and 6-311G(d,p), whereas dipole moments are better described by PBE0 and $\omega$B97X-D3 with aug-cc-pVTZ and aug-cc-pVQZ. Ionisation potentials show a stronger functional dependence, with M06-2X and $\omega$B97X-D3 giving the closest values to experiment, while polarisability is mainly controlled by the inclusion of diffuse functions and sufficient basis-set flexibility. Overall, $\omega$B97X-D3 with the aug-cc basis sets provides the most balanced compromise across the target properties. The close agreement between aug-cc-pVTZ and aug-cc-pVQZ also suggests that aug-cc-pVTZ is sufficient for reliable geometry optimisation, while aug-cc-pVQZ provides a more complete target description for the subsequent R-matrix calculations.

\subsection{Total Cross Section}

\begin{figure}[ht]
\centering
  \includegraphics[scale=0.395]{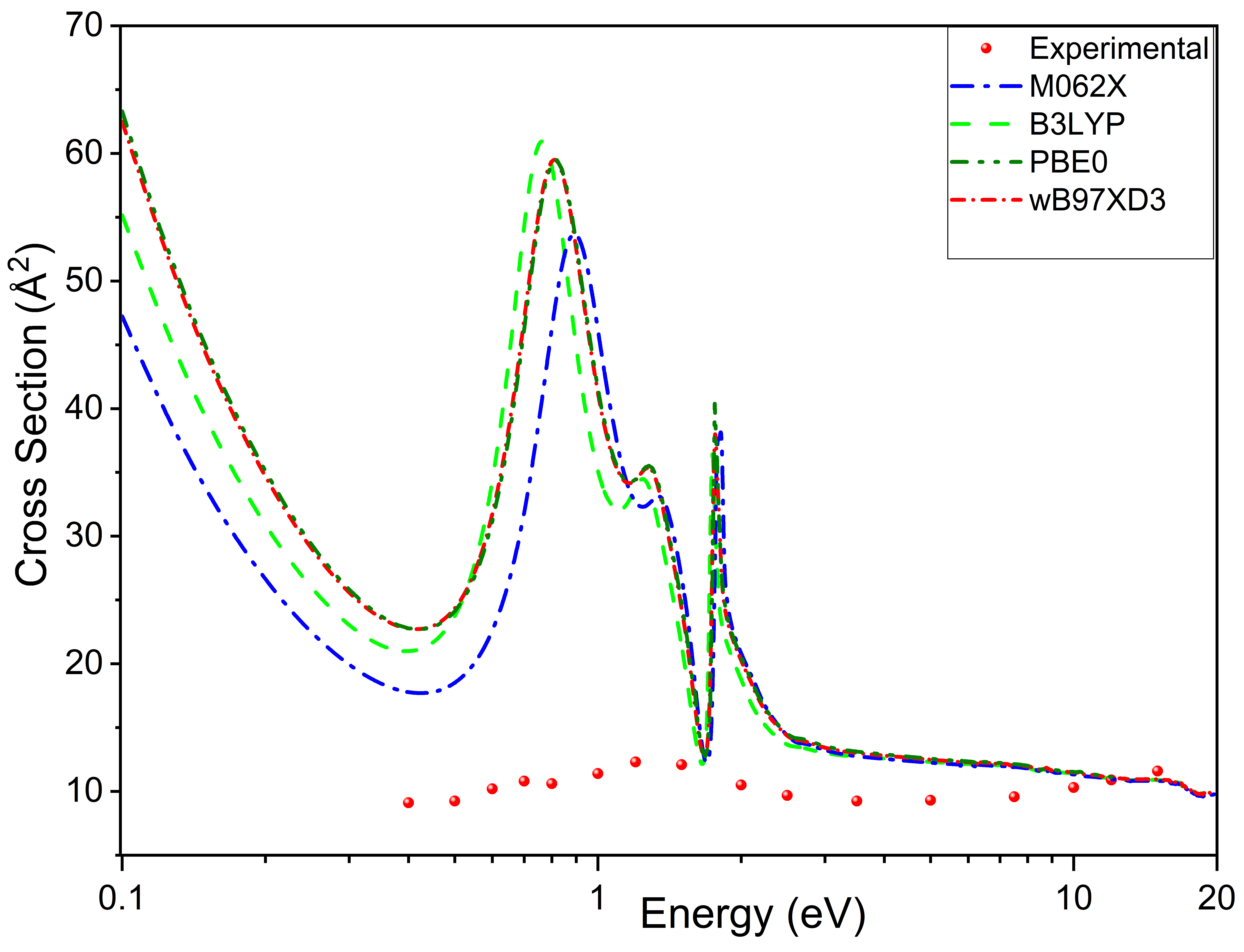}
  \caption{Total electron-scattering cross sections of NO calculated using the R-matrix method with aug-cc-pVQZ target properties obtained from different density functionals. Experimental reference data are shown for comparison.}
  \label{fig5}
\end{figure}

Figure \ref{fig5} shows the total cross sections calculated using the R-matrix method for the NO molecule in the energy range 0.1--20 eV. The CI model described in Section 2 was employed, using molecular properties obtained with the aug-cc-pVQZ basis set and the DFT functionals M06-2X, B3LYP, PBE0, and $\omega$B97X-D3. The corresponding target properties are summarised in Table \ref{tab:target_properties}, together with the available experimental values. This comparison allows the sensitivity of the calculated cross sections to the underlying electronic-structure description to be assessed.

\begin{table}[ht]
\centering
\small
\setlength{\tabcolsep}{5pt}
\caption{Target properties of NO obtained with the aug-cc-pVQZ basis set using different density functionals, compared with experimental values.}
\label{tab:target_properties}
\begin{tabular}{lcccc}
\toprule
Method & 
Bond length(\AA) & 
IP(eV) & 
Dipole moment(D) & 
Polarisability(\AA$^3$) \\
\midrule
Experiment & 1.1508 & 9.264 & 0.159 & 1.698 \\
B3LYP & 1.1445 & 9.580 & 0.142 & 1.567 \\
M06-2X & 1.1356 & 9.405 & 0.133 & 1.689 \\
PBE0 & 1.1392 & 9.539 & 0.154 & 1.709 \\
$\omega$B97X-D3 & 1.1392 & 9.447 & 0.153 & 1.711 \\
\bottomrule
\end{tabular}
\end{table}

At very low energies (0.1--0.4 eV), the cross sections show a clear dependence on the functional. PBE0 and $\omega$B97X-D3 produce nearly overlapping and relatively larger values, while B3LYP gives slightly smaller cross sections and M06-2X yields consistently lower values. 

The most pronounced functional dependence is observed around the first broad resonance peak, centred approximately between 0.8 and 1.0 eV. In this region, the peak position and intensity vary noticeably with the functional. B3LYP predicts the highest and slightly lower-energy peak, PBE0 and $\omega$B97X-D3 give very similar resonance profiles, and M06-2X produces a lower peak shifted to higher energy. This indicates that the first resonance feature is particularly sensitive to the molecular target description. Between this broad peak and the sharper higher-energy resonance, the calculated cross sections also display a shoulder around 1.1--1.3 eV. The persistence of this feature across all functionals suggests that it is associated with overlapping resonance contributions or with interference between resonant and background scattering channels, rather than with a functional-specific numerical effect. Resonance is emphasized in this study because resonances correspond to energies where the interaction probability becomes very large due to the formation of a temporary quasi-bound state during the collision process. A more definitive assignment would require analysis of the eigenphase sums or the symmetry-resolved resonance contributions.

A second, sharper structure is observed at higher energy, with resonance positions around 1.74 eV for B3LYP, 1.76 eV for PBE0 and $\omega$B97X-D3, and 1.82 eV for M06-2X. These values are consistent with experimental observations placing the resonance below 2 eV \citep{RES}. Although this feature also shifts with the functional, the largest functional effect in the calculated total cross sections is associated with the first broad resonance peak.

Beyond the resonance region, the cross sections gradually converge. Between approximately 2 and 10 eV, the functional dependence becomes less significant, except near residual resonance features. Above 10 eV, all curves overlap closely, indicating that the total cross section becomes largely insensitive to the choice of functional at higher collision energies.

Comparison with reference data \citep{NOREF} shows that the calculated cross sections are generally overestimated at low energies, particularly in the resonance region, while better agreement is achieved at higher energies, where the curves approach the experimental values. These results show that the choice of functional has a significant impact on the low-energy resonance structure, especially the first broad resonance peak near 0.8--1.0 eV, whereas the total cross section becomes less sensitive to the electronic-structure method at higher energies.

\subsection{Differential Cross Section}

\begin{figure}[ht]
\centering

\begin{subfigure}{0.49\textwidth}
    \centering
    \includegraphics[width=\linewidth]{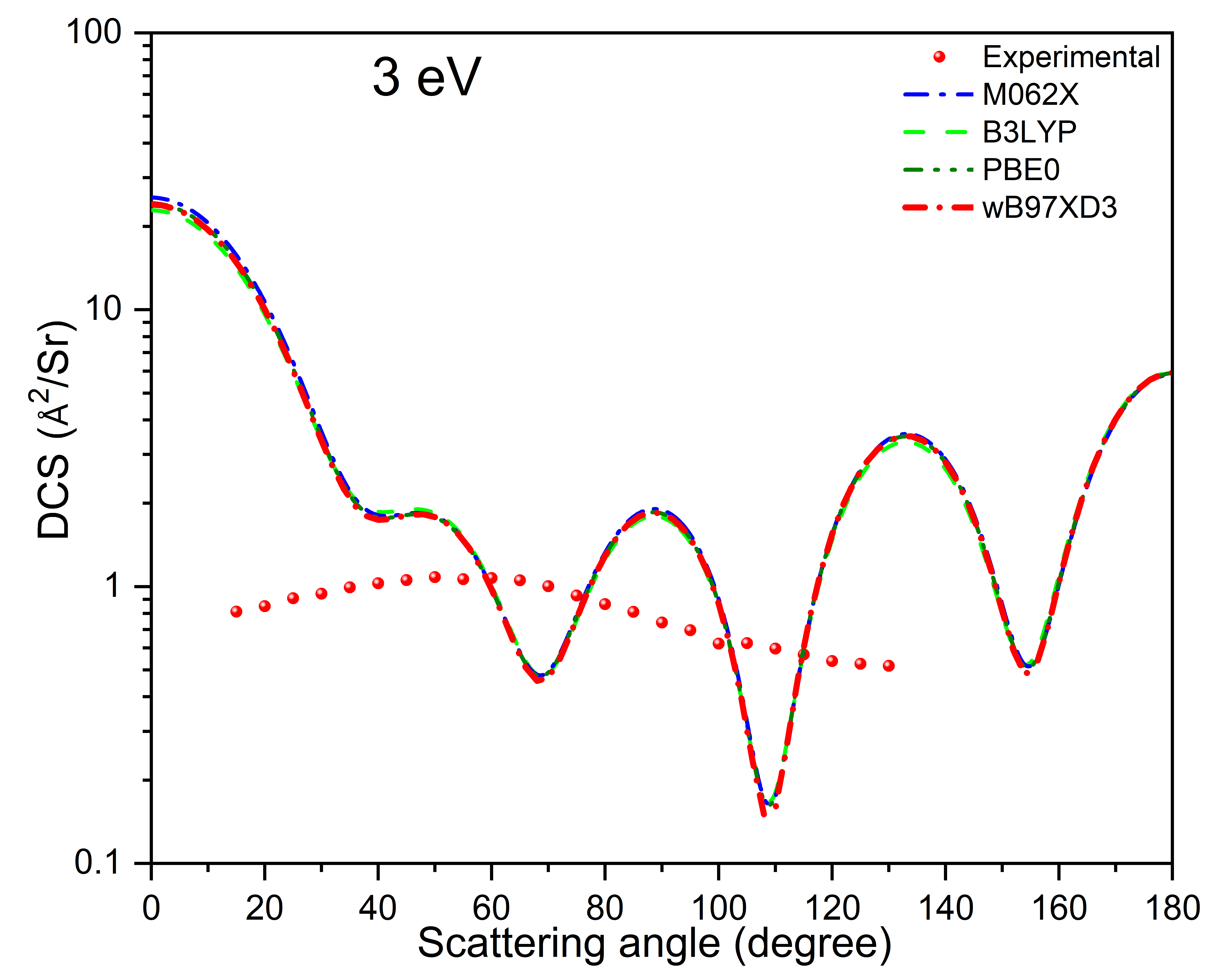}
    \caption{}
\end{subfigure}
\hfill
\begin{subfigure}{0.49\textwidth}
    \centering
    \includegraphics[width=\linewidth]{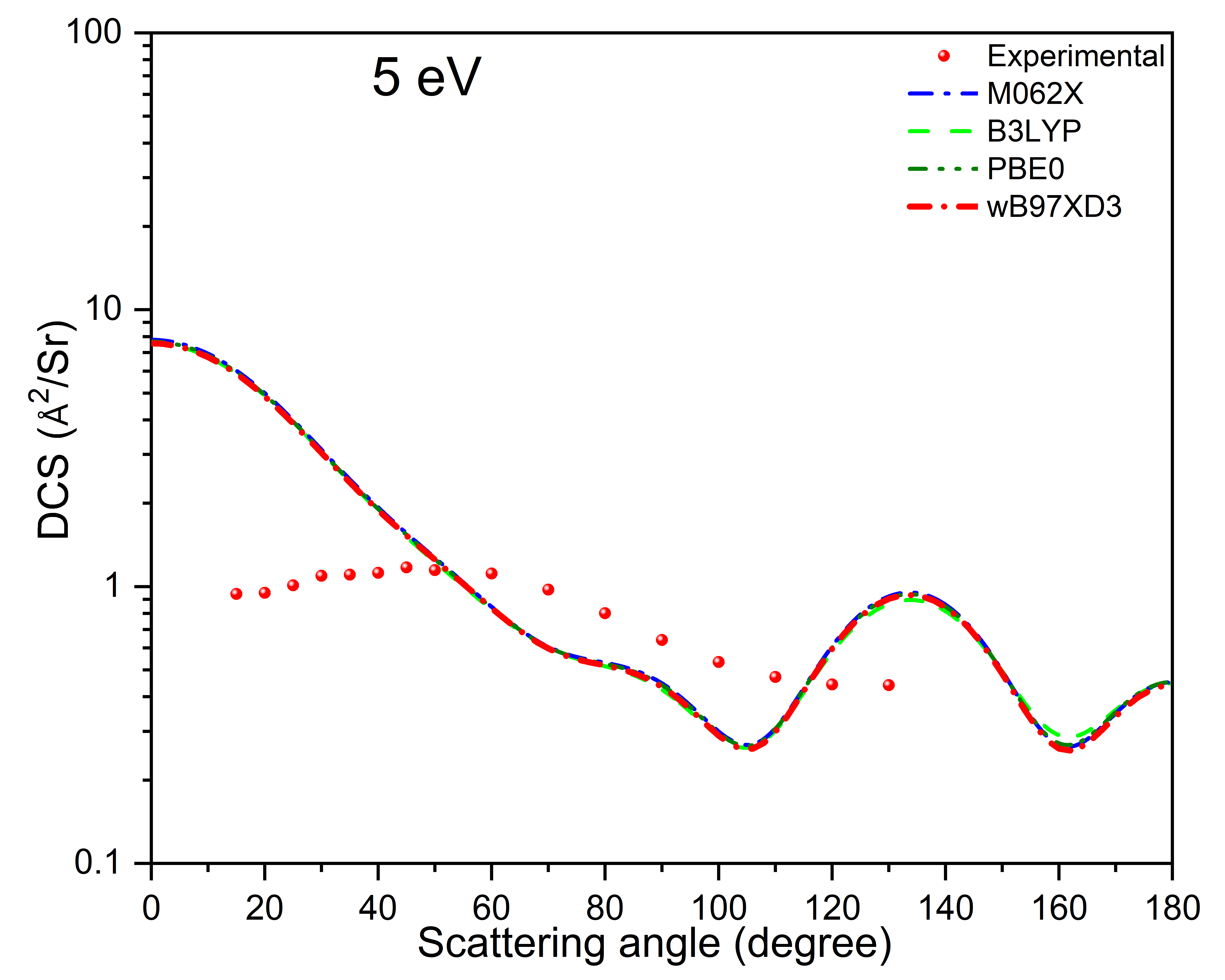}
    \caption{}
\end{subfigure}

\vspace{0.5cm}

\begin{subfigure}{0.49\textwidth}
    \centering
    \includegraphics[width=\linewidth]{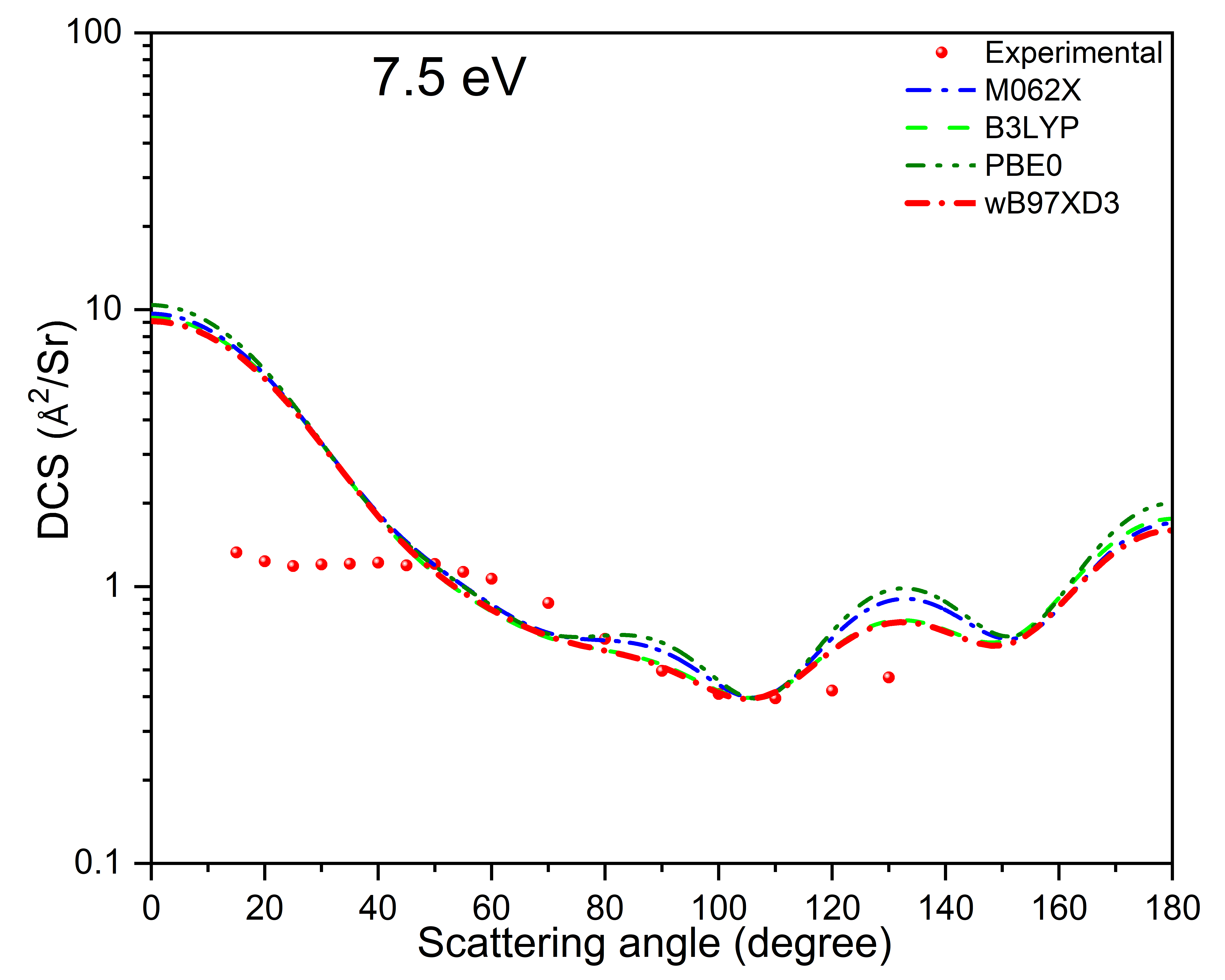}
    \caption{}
\end{subfigure}
\hfill
\begin{subfigure}{0.49\textwidth}
    \centering
    \includegraphics[width=\linewidth]{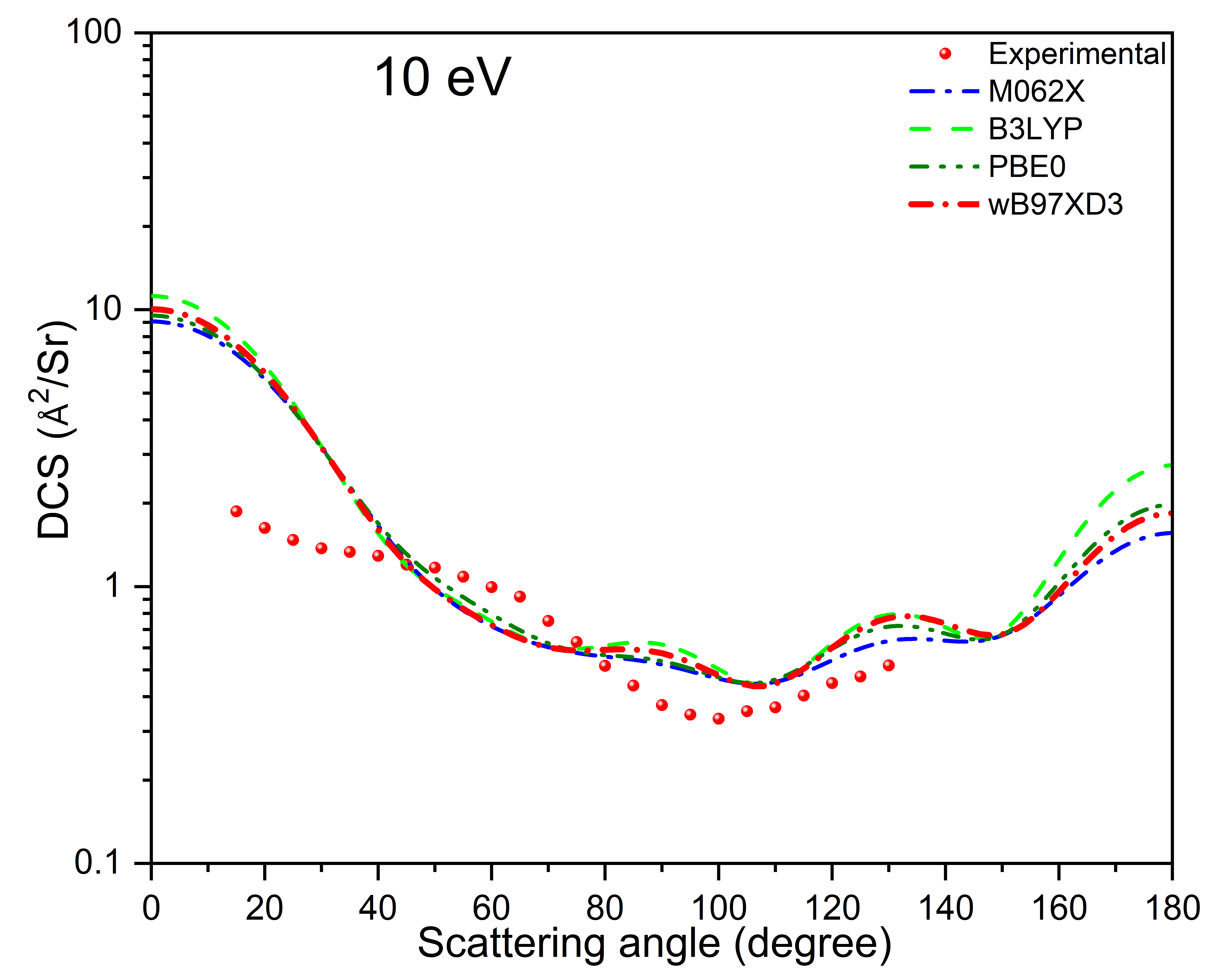}
    \caption{}
\end{subfigure}

\caption{Differential electron-scattering cross sections of NO calculated using the R-matrix method with aug-cc-pVQZ target properties obtained from different density functionals at incident energies of (a) 3 eV, (b) 5 eV, (c) 7.5 eV, and (d) 10 eV. Experimental reference data are shown for comparison.}
\label{DCS}
\end{figure}

The elastic differential cross sections (DCSs) are presented in Figure \ref{DCS}(a--d) for the four density functionals using the aug-cc-pVQZ basis set, at incident electron energies of 3, 5, 7.5, and 10 eV. These calculations were performed to assess the sensitivity of the angular scattering distribution to the choice of functional.

At 3 eV, the DCSs obtained with the different functionals are almost indistinguishable over the full angular range from \(0^\circ\) to \(180^\circ\). The curves show the same shape and magnitude, indicating that the angular distribution is essentially insensitive to the target description at this energy. The calculated DCS is dominated by strong forward scattering, followed by a sequence of minima and maxima at intermediate and backward angles.

A similar trend is observed at 5 eV, where the four calculated curves again follow nearly the same angular profile. Only a slight spread is observed at large scattering angles, particularly around \(160^\circ\). The DCS decreases from the forward direction towards intermediate angles, reaches a broad minimum near \(100^\circ\), and then increases slightly towards backward scattering angles.

\newpage 

At 7.5 eV, the functional dependence becomes more visible. The overall angular profile remains similar for all functionals, but small differences appear in the intermediate- and backward-angle regions. The B3LYP and $\omega$B97X-D3 curves remain close to each other and generally give slightly lower values than M06-2X and PBE0 in parts of the angular range. In the \(90^\circ\)--\(110^\circ\) region, the calculated DCSs show good agreement with the experimental data \citep{NOREF}, although the forward-scattering intensity remains overestimated.

At 10 eV, the DCSs show more noticeable functional dependence in both shape and magnitude. The differences are small over some angular intervals, but become clearer at low scattering angles, below approximately \(20^\circ\), and at larger angles above \(100^\circ\). In the backward-scattering region, B3LYP gives the largest values, while M06-2X generally gives lower values, with PBE0 and $\omega$B97X-D3 lying between them.

The DCSs are less affected by the choice of functional than the total cross sections, particularly in comparison with the strong functional sensitivity observed for the low-energy resonance features. Functional effects in the DCS become more apparent at higher incident energies and at larger scattering angles, but the general angular profiles remain similar across the four electronic-structure descriptions. 

Combining the molecular-property benchmarks with the scattering results, $\omega$B97X-D3/aug-cc-pVQZ provides the most balanced target description for R-matrix modelling of electron scattering from NO. This combination gives a good compromise for the dipole moment, ionisation potential, and polarisability, avoids the larger structural deviation observed for M06-2X, and produces total and differential cross sections consistent with the other well-performing target descriptions. For NO, full $\omega$B97X-D3/aug-cc-pVQZ geometry optimisation and target-property calculations therefore provide a robust reference protocol. As a practical alternative, the close aug-cc-pVTZ/aug-cc-pVQZ agreement supports $\omega$B97X-D3/aug-cc-pVTZ geometry optimisation followed by aug-cc-pVQZ target-property calculations for the R-matrix model.

\section{Conclusion}

In this work, the sensitivity of low-energy electron scattering from nitric oxide (NO) to the underlying DFT-based target description was investigated. The molecular properties of NO, including bond length, dipole moment, ionisation potential, and polarisability, were calculated using four density functionals, B3LYP, M06-2X, PBE0, and $\omega$B97X-D3, which differ in the amount and treatment of Hartree--Fock exchange. These functionals were combined with basis sets ranging from minimal descriptions, such as STO-3G and 3-21G, to triple- and quadruple-zeta basis sets, including aug-cc-pVTZ and aug-cc-pVQZ. The calculated molecular properties were compared with experimental reference data to assess how the choice of functional and basis set affects the target description. The results show that the accuracy of the target properties depends on both the basis set and the exchange--correlation functional, although the relative importance of these factors varies with the property considered. For the bond length, the 6-311++G(d,p) and 6-311G(d,p) basis sets provide the closest agreement with experiment, particularly with B3LYP. The dipole moment is better described by aug-cc-pVTZ and aug-cc-pVQZ in combination with PBE0 and $\omega$B97X-D3, highlighting the importance of diffuse functions for describing charge separation in NO. The ionisation potentials show a clearer dependence on the functional, with M06-2X and $\omega$B97X-D3 giving values closer to experiment, consistent with the role of exact exchange in reducing self-interaction effects. In contrast, the polarisability is mainly controlled by the basis set, with diffuse and sufficiently flexible triple- and quadruple-zeta basis sets giving the best agreement with experiment. 

Using the aug-cc-pVQZ basis set, total and differential cross sections were then calculated with the R-matrix method for target descriptions generated from the different functionals. The total cross sections show marked functional sensitivity in the low-energy region, especially around the first broad resonance peak near 0.8--1.0 eV. A shoulder between the main resonance features is also observed in the computed curves, suggesting overlapping resonance contributions or interference between resonant and background scattering channels. At higher energies, particularly above 10 eV, the total cross sections obtained with different functionals converge and become much less sensitive to the electronic-structure description. 

The differential cross sections (DCS) show weaker functional dependence than the total cross sections. At 3 and 5 eV, the angular distributions are nearly identical for all functionals, while more visible differences appear at 7.5 and 10 eV, especially at larger scattering angles. The results demonstrate that the choice of DFT functional and basis set can significantly affect the molecular target properties and, consequently, the calculated low-energy electron-scattering observables. This effect is most pronounced for resonance features in the total cross section, while the DCS is comparatively less sensitive to the functional choice. Taken together, the comparison supports $\omega$B97X-D3/aug-cc-pVQZ as the most balanced target description for R-matrix modelling of electron scattering from NO. Since aug-cc-pVTZ and aug-cc-pVQZ give closely similar molecular properties, $\omega$B97X-D3/aug-cc-pVTZ geometry optimisation followed by aug-cc-pVQZ target-property calculations also provides a practical alternative protocol.

\appendix


\newpage
\section*{Acknowledgment}

BA and NJM acknowledge the support of the Royal Society, London, for the
International Exchange Award (IES{\textbackslash}R3{\textbackslash}223221) under which this
work was initiated.

\bibliographystyle{elsarticle-harv} 
\bibliography{cas-refs}

@article{Pimblott2007,
author = {Pimblott, Simon M. and LaVerne, Jay A.},
issn = {0969806X},
journal = {Radiation Physics and Chemistry},
month = {aug},
number = {8-9},
pages = {1244--1247},
title = {{Production of low-energy electrons by ionizing radiation}},
volume = {76},
year = {2007},
}

@article{Alizadeh2015,
author = {Alizadeh, Elahe and Orlando, Thomas M. and Sanche, L{\'{e}}on},
journal = {Annual Review of Physical Chemistry},
month = {apr},
number = {1},
pages = {379--398},
title = {{Biomolecular Damage Induced by Ionizing Radiation: The Direct and Indirect Effects of Low-Energy Electrons on DNA}},
volume = {66},
year = {2015},
}

@article{Bogaerts2002,
author = {Bogaerts, Annemie and Neyts, Erik and Gijbels, Renaat and van der Mullen, Joost},
issn = {05848547},
journal = {Spectrochimica Acta Part B: Atomic Spectroscopy},
month = {apr},
number = {4},
pages = {609--658},
title = {{Gas discharge plasmas and their applications}},
volume = {57},
year = {2002},
}

@article{Brunger2017,
author = {Brunger, M. J.},
journal = {International Reviews in Physical Chemistry},
month = {apr},
number = {2},
pages = {333--376},
title = {{Electron scattering and transport in biofuels, biomolecules and biomass fragments}},
volume = {36},
year = {2017},
}

@article{Campbell2016,
author = {Campbell, L. and Brunger, M.J.},
journal = {International Reviews in Physical Chemistry},
month = {apr},
number = {2},
pages = {297--351},
title = {{Electron collisions in atmospheres}},
volume = {35},
year = {2016},
}

@article{Sanche2003,
author = {Sanche, L{\'{e}}on},
issn = {0168583X},
journal = {Nuclear Instruments and Methods in Physics Research Section B: Beam Interactions with Materials and Atoms},
month = {aug},
pages = {4--10},
title = {{Irradiation of organic and polymer films with low-energy electrons}},
volume = {208},
year = {2003},
}

@article{Dickers2025,
author = {Dickers, Matthew D. and Mifsud, Duncan V. and Mason, Nigel J. and Fantuzzi, Felipe},
issn = {0038-6308},
journal = {Space Science Reviews},
month = {dec},
number = {8},
pages = {106},
title = {{Multiscale Perspectives on Solid-Phase Astrochemistry: Laboratory, Computation, and Open Questions}},
volume = {221},
year = {2025}
}

@incollection{Mason2026,
author    = {Mason, Nigel J. and Ioppolo, Sergio and Quiti{\'a}n-Lara, Heidy M. and Fantuzzi, Felipe and Mifsud, Duncan V. and Bockov{\'a}, Jana},
title     = {Foundations of Astrochemistry I: Observations and Ion- and Electron-Induced Processes in Ices},
booktitle = {Advances in Atomic and Molecular Physics at the Interfaces with Natural Sciences, Technology and Medicine},
pages     = {575--621},
publisher = {World Scientific},
year      = {2026}
}

@article{Zhang2024,
author = {Zhang, Jin and Mui{\~{n}}a, Alejandra Traspas and Mifsud, Duncan V and Kaňuchov{\'{a}}, Zuzana and Cielinska, Klaudia and Herczku, P{\'{e}}ter and Rahul, K K and Kov{\'{a}}cs, S{\'{a}}ndor T S and R{\'{a}}cz, Rich{\'{a}}rd and Santos, Julia C and Hopkinson, Alfred T and Craciunescu, Luca and Jones, Nykola C and Hoffmann, S{\o}ren V and Biri, S{\'{a}}ndor and Vajda, Istv{\'{a}}n and Rajta, Istv{\'{a}}n and Dawes, Anita and Sivaraman, Bhalamurugan and Juh{\'{a}}sz, Zolt{\'{a}}n and Sulik, B{\'{e}}la and Linnartz, Harold and Hornek{\ae}r, Liv and Fantuzzi, Felipe and Mason, Nigel J and Ioppolo, Sergio},
journal = {Monthly Notices of the Royal Astronomical Society},
month = {aug},
number = {1},
pages = {826--840},
title = {{A systematic IR and VUV spectroscopic investigation of ion, electron, and thermally processed ethanolamine ice}},
volume = {533},
year = {2024},
}

@article{ARYA2,
author = {Arya, Sudhanshu and Antony, Bobby},
year = {2026},
month = {03},
pages = {13548-13558},
title = {Improved electron-molecule scattering calculations with the relativistic optical-potential method},
volume = {16},
journal = {RSC Advances},
}

@article{NOREF,
author = {Song, Mi-Young and Yoon, Jung-Sik and Cho, Hyuck and Karwasz, Grzegorz and Kokoouline, Viatcheslav and Nakamura, Yoshiharu and Tennyson, Jonathan},
year = {2019},
month = {12},
pages = {043104},
title = {Cross Sections for Electron Collisions with {NO}, {N$_2$O}, and {NO$_2$}},
volume = {48},
journal = {Journal of Physical and Chemical Reference Data},
}

@article{DEANO,
author = {Lozano, Ana I. and Oller, Juan C. and Limão-Vieira, Paulo and García, Gustavo},
title = {Electron Attachment to Nitric Oxide (NO) Controversy},
journal = {The Journal of Physical Chemistry A},
volume = {129},
number = {10},
pages = {2429-2433},
year = {2025},
}

@article{HT,
author = {Tomer, Himani and Goswami, Biplab and Modak, Paresh and Alam, Mohammad and Ahmad, Shabbir and Antony, Bobby},
year = {2023},
month = {12},
pages = {10464–10480},
title = {Low-Energy Electron Scattering from Pyrrole and Its Isomers},
volume = {127},
journal = {The Journal of Physical Chemistry A},
}

@article{PM,
author = {Modak, Paresh and Singh, Abhisek and Goswami, Biplab and Antony, Bobby},
year = {2021},
month = {10},
pages = {},
title = {Electron collision with N$_2$H and HCO},
volume = {75},
journal = {The European Physical Journal D},
}

@article{NAGHMA201317,
title = {Total and elastic cross sections for methyl halides by electron impact},
author = {Rahla Naghma and Bobby Antony},
journal = {Journal of Electron Spectroscopy and Related Phenomena},
volume = {189},
pages = {17-22},
year = {2013},
}

@article{AY2025,
title={Investigation of Electron Collisions with Organic Phosphates},
author={Yadav, Ashutosh and Antony, Bobby},
journal={The Journal of Physical Chemistry A},
volume={129},
pages={8013--8023},
year={2025},  
}

@article{30YEARSDFT,
author = {Narbe Mardirossian and Martin Head-Gordon},
title = {Thirty years of density functional theory in computational chemistry: an overview and extensive assessment of 200 density functionals},
journal = {Molecular Physics},
volume = {115},
number = {19},
pages = {2315--2372},
year = {2017},
}

@MISC{CCCBDB,
  title={Computational Chemistry Comparison and Benchmark Database},
  author={NIST},
  year={2025},
  url={https://cccbdb.nist.gov/},
}

@article{Shiotari2021,
author = {Shiotari, Akitoshi and Koshida, Hiroyuki and Okuyama, Hiroshi},
journal = {Surface Science Reports},
month = {mar},
number = {1},
pages = {100500},
title = {{Adsorption and valence electronic states of nitric oxide on metal surfaces}},
volume = {76},
year = {2021},
}

@article{Barth2003,
author = {Barth, C. A. and Mankoff, K. D. and Bailey, S. M. and Solomon, S. C.},
issn = {0148-0227},
journal = {Journal of Geophysical Research: Space Physics},
month = {jan},
number = {A1},
title = {{Global observations of nitric oxide in the thermosphere}},
volume = {108},
year = {2003}
}

@article{B3LYP,
author = {Becke, Axel D.},
title = {Density‐functional thermochemistry. III. The role of exact exchange },
journal = {The Journal of Chemical Physics},
volume = {98},
number = {7},
pages = {5648-5652},
year = {1993},
month = {04}
}

@article{Krasnopolsky2006,
author = {Krasnopolsky, Vladimir A.},
issn = {00191035},
journal = {Icarus},
month = {may},
number = {1},
pages = {80--91},
title = {{A sensitive search for nitric oxide in the lower atmospheres of Venus and Mars: Detection on Venus and upper limit for Mars}},
volume = {182},
year = {2006}
}

@article{Stoffels2006,
author = {Stoffels, E and Gonzalvo, Y Aranda and Whitmore, T D and Seymour, D L and Rees, J A},
issn = {0963-0252},
journal = {Plasma Sources Science and Technology},
month = {aug},
number = {3},
pages = {501--506},
title = {{A plasma needle generates nitric oxide}},
volume = {15},
year = {2006}
}

@article{Liu2025,
author = {Liu, Xiangtao and Si, Jicang and Wang, Guochang and Wu, Mengwei and Mi, Jianchun},
issn = {03605442},
journal = {Energy},
month = {jan},
pages = {134455},
title = {{Nitrogen sources and formation routes of nitric oxide from pure ammonia combustion}},
volume = {315},
year = {2025},
}

@article{PBE0,
author = {Adamo, Carlo and Barone, Vincenzo},
journal = {The Journal of Chemical Physics},
month = {apr},
number = {13},
pages = {6158--6170},
title = {Toward reliable density functional methods without adjustable parameters: The PBE0 model},
volume = {110},
year = {1999},
}

@article{M06,
author = {Zhao, Yan and Truhlar, Donald},
year = {2008},
month = {04},
pages = {215-241},
title = {The {M06} suite of density functionals for main group thermochemistry, thermochemical kinetics, noncovalent interactions, excited states, and transition elements: two new functionals and systematic testing of four {M06} functionals and 12 other functionals},
volume = {120},
journal = {Theoretical Chemistry Accounts},
}

@article{tennyson1,
  title={Electron--molecule collision calculations using the {R}-matrix method},
  author={Tennyson, Jonathan},
  journal={Physics Reports},
  volume={491},
  number={2-3},
  pages={29--76},
  year={2010},
}

@inproceedings{tennyson2,
  title={Quantemol-N: an expert system for performing electron molecule collision calculations using the R-matrix method},
  author={Tennyson, Jonathan and Brown, Daniel B and Munro, James J and Rozum, Iryna and Varambhia, Hemal N and Vinci, Natalia},
  booktitle={Journal of Physics: Conference Series},
  volume={86},
  number={1},
  pages={012001},
  year={2007},
}

@article{RES,
author = {Itikawa, Yukikazu},
    title = {Cross sections for electron collisions with nitric oxide},
    journal = {Journal of Physical and Chemical Reference Data},
    volume = {45},
    number = {3},
    pages = {033106},
    year = {2016},
    month = {09},
}

@article{gailitis,
  title={New forms of asymptotic expansions for wavefunctions of charged-particle scattering},
  author={Gailitis, M},
  journal={Journal of Physics B: Atomic and Molecular Physics},
  volume={9},
  number={5},
  pages={843},
  year={1976},
}

@article{sanna1998differential,
  title={Differential cross sections for electron/positron scattering from polyatomic molecules},
  author={Sanna, Nico and Gianturco, FA},
  journal={Computer physics communications},
  volume={114},
  number={1-3},
  pages={142--167},
  year={1998},
}

@article{neese2020orca,
author={Neese, Frank and Wennmohs, Frank and Becker, Ute and Riplinger, Christoph},
title={The \textsc{ORCA} quantum chemistry program package},
journal={The Journal of Chemical Physics},
year={2020},
volume={152},
pages = {224108},
}

@incollection{Hughes2008,
title = {Chapter One - Chemistry of Nitric Oxide and Related Species},
editor = {Robert K. Poole},
series = {Methods in Enzymology},
publisher = {Academic Press},
volume = {436},
pages = {3-19},
year = {2008},
booktitle = {Globins and Other Nitric Oxide-Reactive Proteins, Part A},
issn = {0076-6879},
author = {Martin N. Hughes},
}

@article{Lin2013,
author = {Lin, You-Sheng and Li, Guan-De and Mao, Shan-Ping and Chai, Jeng-Da},
issn = {1549-9618},
journal = {Journal of Chemical Theory and Computation},
month = {jan},
number = {1},
pages = {263--272},
title = {{Long-Range Corrected Hybrid Density Functionals with Improved Dispersion Corrections}},
volume = {9},
year = {2013}
}

@article{Liszt1978,
author = {Liszt, H. S. and Turner, B. E.},
issn = {0004-637X},
journal = {The Astrophysical Journal},
month = {sep},
pages = {L73},
title = {{Microwave detection of interstellar NO}},
volume = {224},
year = {1978},
}

@article{Ernzerhof1999,
author = {Ernzerhof, Matthias and Scuseria, Gustavo E.},
issn = {0021-9606},
journal = {The Journal of Chemical Physics},
month = {mar},
number = {11},
pages = {5029--5036},
title = {{Assessment of the Perdew–Burke–Ernzerhof exchange-correlation functional}},
volume = {110},
year = {1999},
}

@article{Huh2010,
author = {Huh, Do Sung and Choe, Sang Joon},
issn = {1088-4246},
journal = {Journal of Porphyrins and Phthalocyanines},
month = {jul},
number = {07},
pages = {592--604},
title = {{Comparative DFT study for molecular geometries and spectra of methyl pheophorbides-a: test of M06-2X and two other functionals}},
volume = {14},
year = {2010}
}

@article{Bao2018,
author = {Bao, Junwei Lucas and Gagliardi, Laura and Truhlar, Donald G.},
issn = {1948-7185},
journal = {The Journal of Physical Chemistry Letters},
month = {may},
number = {9},
pages = {2353--2358},
title = {{Self-Interaction Error in Density Functional Theory: An Appraisal}},
volume = {9},
year = {2018}
}

@article{Zhao2008,
author = {Zhao, Yan and Truhlar, Donald G.},
issn = {1549-9618},
journal = {Journal of Chemical Theory and Computation},
month = {nov},
number = {11},
pages = {1849--1868},
title = {{Exploring the Limit of Accuracy of the Global Hybrid Meta Density Functional for Main-Group Thermochemistry, Kinetics, and Noncovalent Interactions}},
volume = {4},
year = {2008}
}

@article{Kirschner2020,
author = {Kirschner, Karl N. and Reith, Dirk and Heiden, Wolfgang},
issn = {1539-445X},
journal = {Soft Materials},
month = {jul},
number = {2-3},
pages = {200--214},
title = {{The performance of Dunning, Jensen, and Karlsruhe basis sets on computing relative energies and geometries}},
volume = {18},
year = {2020}
}





\end{document}